\documentclass[journal]{IEEEtran}
\usepackage{cite}

\usepackage{amsmath,amssymb,fancyhdr,wasysym,graphicx}
\usepackage{graphicx} 
\usepackage{multirow} 
\usepackage{amsfonts}
\usepackage{url}
\usepackage{xcolor}
\usepackage{threeparttable}
\usepackage[perpage,symbol]{footmisc}
\usepackage{booktabs}
\usepackage{makecell}

\setfnsymbol{wiley}

\usepackage{CJK}
\ifCLASSINFOpdf
\else
\fi
\begin{document}

\title{Set-based Obfuscation for Strong PUFs against Machine Learning Attacks}
%
%
%



\author{Jiliang~Zhang~\IEEEmembership{Senior Member,~IEEE}, Chaoqun Shen
\thanks{This work was supported by the National Natural Science Foundation of China under Grant No. 61874042 and 61602107, the Hu-Xiang Youth Talent Program under Grant No. 2018RS3041, the Key Research and Development Program of Hunan Province under Grant No. 2019GK2082, and the Fundamental Research Funds for the Central Universities. (\textit{Corresponding author: Jiliang Zhang}}

\thanks{J. Zhang and C. Shen is with the College of Computer Science and Electronic Engineering, Hunan University, Changsha 410082, China (e-mail: zhangjiliang@hnu.edu.cn).}

}
\maketitle

\begin{abstract}
Strong physical unclonable function (PUF) is a promising solution for device authentication in resource-constrained applications but vulnerable to machine learning attacks.  In order to resist such attack, many defenses have been proposed in recent years. However, these defenses incur high hardware overhead, degenerate reliability and are inefficient against advanced machine learning attacks such as approximation attacks. In order to address these issues, we propose a \underline{R}andom \underline{S}et-based \underline{O}bfuscation (RSO) for Strong PUFs to resist machine learning attacks. The basic idea is that several stable responses are derived from the PUF itself and pre-stored as the $set$ for obfuscation in the testing phase, and then a true random number generator is used to select any two keys to obfuscate challenges and responses with XOR operations. When the number of challenge-response pairs (CRPs) collected by the attacker exceeds the given threshold, the $set$ will be updated immediately. In this way, machine learning attacks can be prevented with extremely low hardware overhead. Experimental results show that for a 64$\times$64 Arbiter PUF, when the size of $set$ is 32 and even if 1 million CRPs are collected by attackers, the prediction accuracies of Logistic regression, support vector machines, artificial neural network, convolutional neural network and covariance matrix adaptive evolutionary strategy are about 50$\%$ which is equivalent to the random guessing.

\end{abstract}


%
\IEEEpeerreviewmaketitle

\section{Introduction}
\subsection{Background and Motivation}
Internet of things (IoT) is the network of physical devices, vehicles, home appliances and other items embedded with electronics, software, sensors, actuators, and the connectivity enables these objects to connect and exchange data. According to IHS forecast \cite{iot2018}, the global installed IoT devices will increase from 15.4 billion in 2015 to 30.7 billion in 2020, and this number will reach 75.4 billion in 2025. With the rapid development of IoT, security issues attracted much attention and became particularly serious. In 2018, the IoT Security White Paper \cite{iotwp2018} pointed out that the global total cost of IoT security was 1174 million dollars in 2017, reached 1506 million dollars in 2018 and was estimated at about 1931 million dollars in 2019. Security issues have govern the sustainable development of the IoT. Secret key storage and device authentication are the two key technologies to address IoT security issues. Traditional security mechanisms store secret keys in electrically erasable programmable read-only memory (EEPROM) or battery-backed non-volatile static random access memory (SRAM), and implement information encryption and authentication with cryptographic algorithms. However, in many IoT applications, resources like CPU, memory, and battery power are limited so that they cannot afford the classic cryptographic security solutions. Therefore, lightweight solutions for IoT security are urgent.

Physical unclonable function (PUF) is an alternative solution for low cost key generation and device authentication. It is a physical entity that is embodied in a physical structure and easy to manufacture and evaluate but practically impossible to duplicate, even with the exact same manufacturing process \cite{Wikipedia2013}. In the past decade, intensive study has focused on PUFs and lots of PUF structures are proposed such as Arbiter PUF \cite{Gassend2004,Lee2004,Majzoobi2009}, SRAM PUF \cite{Holcomb2007} and ring oscillator (RO) PUF \cite{Suh2007}. These PUFs can be classed into the strong PUF \cite{Lee2004,Sahoo2014,Majzoobi2008a,Vijayakumar2015} and weak PUF \cite{Holcomb2007,Tuyls2006,Sauer2017,Suh2007}. Weak PUFs exhibit only a small number of challenge-response pairs (CRPs) which can be used as a device unique key or seed for conventional encryption systems. On the other hand, strong PUFs such as Arbiter PUF can provide a huge number of unique CRPs, which enables the strong PUFs suitable for lightweight device authentication. However, current strong PUFs are vulnerable to machine learning (ML) attacks that attackers can collect a certain number of CRPs from the communication channel to model (clone) the PUF structure \cite{R¨¹hrmair2010}. For example, for a 64$\times$64 Arbiter PUF, the predication accuracy of trained soft model can reach up to 99.9$\%$ when 18050 pairs of CRPs are used. The cloned soft PUF exhibits almost the same challenge-response behavior as the hardware one.

\subsection{Limitations of Prior Art}
In order to resist ML attacks, many defenses are proposed and can be roughly classed into structural non-linearization and CRP obfuscation. Structural non-linearization methods \cite{Gassend2004,Sahoo2018,Kumar2014,Vijayakumar2015} are to implement the nonlinear mapping of CRPs by designing the specific non-linearization PUF structures. However, the vast majority of existing ML-resistant PUFs reduce the reliability of responses largely and can be modeled with high accuracy \cite{R¨¹hrmair2010,Ruhrmair2013}. CRP obfuscation methods \cite{Suh2007,Majzoobi2008a,Gassend2008,Gao2017,Rostami2014,Ye2017} are to prevent ML attackers from collecting enough effective CRPs to model PUF via obfuscating the mapping of CRPs. However, current CRP obfuscation methods share several weaknesses: 1) vulnerability to advanced ML attacks such as approximation
attacks and CMA-ES \cite{Becker2015,Delvaux2017}; 2) prohibitively expensive obfuscation structures such as hash functions; 3) reducing the reliability of strong PUFs.

\subsection{Our Contributions}
To solve the limitations of prior art, this paper proposes a simple but efficient defense, named  \underline{R}andom \underline{S}et-based \underline{O}bfuscation (RSO). In addition, the RSO-based PUF authentication protocol is proposed. The main contributions of this paper are as follows.
\begin{enumerate}
  \item \textbf{Universality}.
  The RSO and corresponding authentication protocol can be used for all strong PUFs to resist ML attacks.
  \item \textbf{Low overhead}.
  The RSO just uses XOR logic, a true random number generator (TRNG) and a random response $set$ derived from the PUF itself to obfuscate the mapping of CRPs, which incurs negligible hardware overhead.
  \item \textbf{No effects on reliability}.
  PUF challenges and responses are obfuscated with the random numbers in the $set$ by bitwise XOR operation, which will not reduce the reliability of PUFs.
  \item \textbf{Updateable}.
  The RSO selects any two random numbers from the $set$ to obfuscate the map relationship of CRPs. When the number of CRPs collected by attackers exceeds the threshold we preset, the set will be updated immediately. In this way, RSO is able to resist all advanced ML attacks effectively in theory.
  \item \textbf{Efficient}.
  We have evaluated six ML attacks including Logistic regression (LR), support vector machines (SVM), artificial neural network (ANN), convolutional neural network (CNN) and covariance matrix adaptive evolutionary strategy (CMA-ES) and recently proposed approximation attacks \cite{Zhang2020}. Experimental results show that the prediction accuracies are about 50$\%$ even if 1 million CRPs are collected by attackers when the size of $set$ is 32.

\end{enumerate}

The rest of this paper is organized as follows. Related work is elaborated in Section II. Section III introduces some related definitions, concepts and terminologies. Section IV gives a detailed introduction about our proposed obfuscation method. Experimental results and analysis are reported in Section V. In section VI, we compare the RSO with several recent proposed defenses. Finally, a conclusion is made in Section VII.

\section{Related Work}
In 2004, Lee et al. \cite{Lee2004} demonstrated that the ML attack is a great threat to strong PUFs for the first time. Later, several ML attacks have been proposed to model various strong PUFs \cite{R¨¹hrmair2010,Ruhrmair2013}. In order to resist such attack, many defenses have been proposed in recent years. These defenses can be roughly classed into structural non-linearization and CRP obfuscation.
	
\subsection{Structural Non-linearization}
Structural non-linearization is to implement nonlinear PUF structures to obstruct the ML-based modeling. Feed forward Arbiter PUF (FFA PUF) \cite{Gassend2004}, current mirrors PUF \cite{Kumar2014}, and voltage transfer PUF \cite{Vijayakumar2015} are typical non-linearization PUF structures. In the FFA PUF structure, the racing results of the previous multiplexer stages is feed forward to the following one or more select lines (challenge bits), thus making the Arbiter PUF structure non-linear. The current mirror PUF is to transmit the current through identical non-linear current mirrors, while voltage transfer PUF is to implement a non-linear voltage transfer function. These non-linear PUF structures can resist traditional ML attacks such as LR effectively. However, the reliability is reduced due to the non-linearization of PUF structure. In addition, these non-linear PUF structures are vulnerable to ES-based modeling attacks.

\subsection{CRP Obfuscation}
CRP obfuscation can hide the mapping of CRPs to prevent attackers from collecting valid CRPs to model strong PUFs. Some typical obfuscation methods have been proposed to obfuscate challenges and/or responses with XOR gates, hash functions and random bits.

\subsubsection{XOR gate/ Multiplexer}
XOR and Multiplexer are simple and efficient obfuscation against machine leaning attacks. Suh et al. \cite{Suh2007} proposed a XOR Arbiter PUF that the outputs of multiple parallel Arbiter PUF structures are XORed together to generate 1-bit more secure response. XOR Arbiter PUF improves the ability of resisting machining learning because the mapping of CRPs is obfuscated with the XOR gates at the expense of reduced reliability and high hardware overhead. Wang et al. \cite{Wang2018} propose a feedback structure to XOR the PUF response with the challenge, which ignores the reliability of PUF responses. Majzoobi et al. proposed the lightweight secure PUF \cite{Majzoobi2008a} where complex challenge mapping that derives the individual challenges from a global challenge is applied to multiple parallel Arbiter PUF structures and then multiple individual responses are XORed to produce a multi-bit response. Sahoo et al. \cite{Sahoo2018} proposed to obfuscate the arbiter PUF with multiplexers instead of XOR gates, which shows higher reliability and security. Unfortunately, XOR and multiplexer-based obfuscation methods can been broken by recently proposed approximation attacks \cite{Zhang2020}.



\subsubsection{Hash functions}
Controlled PUF \cite{Gassend2008} is to obfuscate the mapping of CRPs with the hash function. Since both challenges and responses are hashed, the real challenges and responses can't be accessed. However, the hash and error-correction code (ECC) blocks introduce significant area and power overhead, and the use of helper data will make the PUF vulnerable to side-channel attacks \cite{Becker2015,Becker2014}. In order to reduce overhead, Gao et al. \cite{Gao2017} proposed a PUF-FSM obfuscation method that removes the hash logic on challenges and replace the ECC unit with FSM. However, the hash logic for responses still incurs large overhead, and the PUF-FSM obfuscation has been broken by the variant CMA-ES \cite{Delvaux2017}.

\subsubsection{Random bits}
Yu et al. proposed a lockdown technique \cite{Yu2016} which uses a query mechanism to restrict the number of available CRPs for attackers to clone PUF. However, the number of CRPs for authentication is limited. Majzoobi et al. \cite{Rostami2014} proposed a Slender PUF that randomly selects a substring of the response and fill it with a random binary string to ensure that its length is the same as the full response, then the server exploits a recovery method to match the substring selected randomly and authenticate the legality. Ye et al. \cite{Ye2017} proposed a RPUF that randomizes challenges with a random number generator (RNG) before inputting to the strong PUF to prevent ML attacks. However, Slender PUF and RPUF are vulnerable to advanced ML attacks such as CMA-ES \cite{Becker2015,Delvaux2017}.

As discussed above, existing structural non-linearization methods and the vast majority of CRP obfuscation methods degenerate the reliability of PUFs. CRP obfuscation incurs prohibitive overhead. Moreover, structural non-linearization and CRP obfuscation are not completely immune to advanced ML attacks. This paper proposes a simple, but very efficient and low overhead obfuscation technique to resist machine leaning attacks without reducing the reliability.

\section{Preliminaries}
This section will introduce some terminologies and concepts used in this paper and some more detailed definitions will be given when necessary. Throughout this paper, we employ the symbols and terminology shown in Table \ref{table1}.

\begin{table}
\caption{List of parameters.}
\centering
\label{table1}
\begin{tabular}{|c|c|}
\hline
\textbf{Symbol}&\textbf{Description}\\
\hline
\hline
\emph{n} & Number of PUF stages \\
\hline
\emph{C}$_{ob}$ &The set of challenges used for obfuscation \\
\hline
\emph{R}$_{ob}$ & The set of responses used for obfuscation \\
\hline
\emph{K} & The set of keys used for obfuscation \\
\hline
$\emph{Key}_i$ &A key in $K$ used to XOR with PUF challenges \\
\hline
$\emph{Key}_j$ & A key in $K$ used to XOR with PUF responses \\
\hline
$\widehat{SPUF}_i$ & The soft PUF model trained in enrollment phase \\
\hline
$\epsilon$ & The prediction error rate of soft PUF model \\
\hline
$N_{min,\epsilon}$& \thead{Minimum number of CRPs required for attackers \\to build a PUF model with a prediction error rate $\epsilon$} \\
\hline
$n_{tolerance}$ & \thead{Maximum number of bit-flips allowed in a PUF response} \\
\hline
$\tau$ & \thead{The authentication threshold used by the server to \\compare the emulated response $R'$ from the server and \\the received response $R$ from the physical PUF} \\
\hline
$P_{suc}$ & \thead{The probability of successfully authenticating a legal \\PUF-embedded device}\\
\hline
\thead{$n_{EER}$} & \thead{The error threshold which makes the FAR and FRR \\approximately equal} \\
\hline
\end{tabular}
\end{table}

\subsection{Notation}
\begin{enumerate}
  \item [1)]\emph{HD, FHD, mean HD}
\end{enumerate}

\textbf{Hamming distance (HD)}. For the \emph{L}-bit binary strings \emph{X} and \emph{Y}, the HD between \emph{X} and \emph{Y} is defined as:
\begin{equation}
HD(X,Y)=\sum_{i=0}^{L-1} {X[i]}\oplus{Y[i]}
\end{equation}

\textbf{Fractional Hamming distance (FHD)}. The Fractional Hamming distance between \emph{X} and \emph{Y} is defined as:
\begin{equation}
FHD(X,Y)=\frac{HD(X,Y)}{L}
\end{equation}

\textbf{Mean of pairwise HD}. Given a set \textbf{\emph{C}} containing multiple binary strings, the mean of pairwise HD of \textbf{\emph{C}} is defined as:
\begin{equation}
mean HD(\textbf{C})=mean\{f_{HD}(C_i,C_j)\}
\end{equation}
where the binary strings \emph{C}$_i$ $\in$ \textbf{\emph{C}}, \emph{C}$_j$ $\in$ \textbf{\emph{C}} and \emph{i}$\not=$\emph{j}.

\begin{enumerate}
  \item [2)]\emph{Inter-HD, Intra-HD, $\hat{P}$$_{inter}$, $\hat{P}$$_{intra}$}
\end{enumerate}

Inter-HD and intra-HD are used to describe the statistical characteristics of PUF responses. The definitions of inter-HD and intra-HD are as follows.

\textbf{Inter-HD}. Inter-HD indicates the HD between the responses generated by two different PUF instances when the same challenge is input. Inter-HD is used to measure the uniqueness of PUF.

\begin{equation}
Inter\text{-}HD = HD(R_{1};R_{2})
\end{equation}
where $R_1$ and $R_2$ are generated by any two different PUF instances when inputting the same challenge.

\textbf{Intra-HD}. Intra-HD indicates the HD between the responses generated by the same PUF instance when the same challenge is input. Intra-HD is used to measure the reliability of PUF.

\begin{equation}
Intra\text{-}HD = HD(R_{X};R_{Y})
\end{equation}
where $R_X$ and $R_Y$ are generated by the same PUF instance when inputting the same challenge in different environments.

Since both inter-HD and intra-HD distributions follow a binomial distribution, $B(n, \hat{p})$, the binomial probability estimator of inter-HD and intra-HD distributions are
\begin{equation}
\hat{P}_{inter} = P\{R_1\neq R_2\}
\end{equation}
\begin{equation}
\hat{P}_{intra} = P\{R_X\neq R_Y\}
\end{equation}
where $\hat{P}$$_{inter}$ is the probability of $R_1\not = R_2$,  $\hat{P}$$_{intra}$ is the probability of $R_X\not = R_Y$.

\subsection{Strong PUFs}
Strong PUFs can generate a large number of CRPs, scaling exponentially with the required IC area \cite{Maes2013}. Arbiter PUF \cite{Lee2004}, lightweight secure PUF \cite{Majzoobi2008a} and current mirror PUF \cite{Vijayakumar2015} are typical strong PUFs. Among them, Arbiter PUF \cite{Lee2004} is the most popular one. The functionality of Arbiter PUF can be represented by an additive linear delay model \cite{R¨¹hrmair2010,Lim2005,Burleson2016}. When modeling an Arbiter PUF, the total delay of the signals is the cumulative sum of the delay in each stage. In this model, we can define the final delay difference $\Delta$ between the upper and the lower path as:

\begin{equation}
\Delta = \vec {\omega}^T\vec{\phi}
\end{equation}
where $\vec {\omega} = \{ {\omega^1,\omega^2,...,\omega^n,\omega^{n+1}}\}$, the dimensions of $\vec {\omega}$ and $\vec{\phi}$ are both $n + 1$; the eigenvector $\vec {\phi}$ represents a function with the $n$-bit challenge \cite{R¨¹hrmair2010}\cite{Lim2005}\cite{Burleson2016}. The parameter vector $\vec {\omega}$ represents the delay of each stage in an Arbiter PUF; $\omega^1$ = $(\sigma_1^0-\sigma_1^1)$, $\omega^i$ = $(\sigma_{i-1}^0+\sigma_{i-1}^1+\sigma_{i}^0-\sigma_i^1)$ , $i = 2, 3, ..., n$ and $\omega^{n+1} = (\sigma_n^0+\sigma_n^1)/2$. $\sigma$$_{i}$$^{0\diagup1}$ denotes the delay of the multiplexer $M_i$, where $\sigma$$_{i}$$^{1}$ means that the signal is crossed in the $M_i$, while $\sigma$$_{i}$$^{0}$ means uncrossed. In addition,

\begin{equation}
\vec{\phi}(\vec C) = (\phi^1(\vec C),...,\phi^n(\vec C),1)^T
\end{equation}
where $\phi^l(\vec C) = \prod_{i = l}^n (1-2C_i)$, $l = 1,..., n$.

The output of Arbiter PUF $t$ is determined by the sign function acting on the total delay difference. And for convenience, we replace $t = 0$ with $t = -1$:

\begin{equation}
t = sgn(\Delta) = sgn(\vec {\omega}^T\vec{\phi})
\end{equation}

Eqn. (10) indicates that the vector $\vec {\omega}$ determines a separate hyperplane in all the eigenvectors by $\vec {\omega}^T\vec{\phi} = 0$. When $t = -1$, all eigenvectors are on one side of the hyperplane. Conversely, when $t = 1$, all eigenvectors are on the other side. Hence, the response of Arbiter PUF can be predicted by the obtained hyperplane.

\subsection{ML Attacks}
\begin{enumerate}
  \item [1)]\emph{Logistic Regression (LR)}
\end{enumerate}

LR \cite{Bishop2015} is a frequently used supervised ML method. When LR is used to model the Arbiter PUF, each challenge \emph{C} = $\{\emph{C}_1, ..., \emph{C}_n\}$ is given a probability \emph{P}(\emph{C}, \emph{t}$|\vec {\omega}$ ) that generates a response $t (t \in \{-1,1\})$. As a technical convention, $t \in \{0,1\}$ is replaced by $t \in \{-1,1\}$. Since the vector $\vec {\omega}$ denotes the delays of the subcomponents (stages) in the Arbiter PUF, the probability $P (C, t|\vec {\omega} )$ is obtained by the logistic regression sigmoid function acting on $f(\vec {\omega})$:

\begin{equation}
\emph{P}(\emph{C},\emph{t}|\vec {\omega} ) = \sigma(tf) = \sigma(1+e^{-tf})^{-1}
\end{equation}

For the training set $M$, the parameter vector $\vec {\omega}$ is adjusted to determine the decision boundary to minimize the negative-likelihood:

\begin{equation}
\widehat{\vec {\omega}} = argmin_{\vec {\omega}}l(M,\vec {\omega}) = argmin_{\vec {\omega}}\sum_{(C,t)\in M}-ln(\sigma(tf(\vec {\omega},C)))
\end{equation}

As there is no suitable way to find the optimal $\vec {\omega}$ directly, the iterative method such as the gradient descent algorithm is used to solve this problem:

\begin{equation}
\nabla l(M,\vec {\omega}) = \sum_{(C,t)\in M}t(\sigma(tf(\vec {\omega},C))-1)\nabla f(\vec {\omega},C)
\end{equation}

We have tested several optimization methods including standard gradient descent, iterative reweighted least squares and Rprop \cite{Bishop2015}\cite{Riedmiller1993}, where RProp gradient descent works best in LR. The classification object of LR is not required to be linearly separable, but the loss function must be differentiable.

\begin{enumerate}
  \item [2)]\emph{Support Vector Machines (SVM)}
\end{enumerate}

SVM \cite{Bishop2015} can perform binary classification and solve the classification tasks by mapping known training instances into a higher-dimensional space. The goal of SVM training is to find the most suitable separation hyperplane and solve the nonlinear classification tasks that cannot be linearly separated in the original space. The separation hyperplane should keep the maximum distance with all vectors of different classifications as much as possible. The vector with the smallest distance to the separation hyperplane is called the support vector. The separation hyperplane is constructed by the two parallel hyperplanes with support vectors of different classifications. The distance between the hyperplanes is called the margin. The key of constructing a good SVM is to maximize the margin while minimizing classification errors and the whole process is regulated by the regularization coefficient $\lambda$.

In well-trained SVMs, kernel functions are often used to solve the problem of support vector selection and classification. There are three frequently-used kernel functions: 1) linear: $K(w,z) = z^T w $ (only solves linearly separable problems);  2) radial basis function (RBF): $K(w,z)$ = $\exp$$((-\|w-z||_2^2)/\sigma^2 )$; 3) multi-layer perception (MLP): $K(w,z) = \tanh$$(\alpha z^T w+\beta)$. Training a good SVM classifier always requires to adjust regularization coefficient $\lambda$, $\sigma^{2}$ (RBF) or $(\alpha,\beta)$ (MLP). In our experiments, we use the SVM with RBF kernel function to model PUF.

\begin{enumerate}
  \item [3)]\emph{Convolutional Neural Network (CNN)}
\end{enumerate}

CNN \cite{Ciresan2012} has been used in image classification widely and achieved great success in graphic recognition such as handwritten digits \cite{Le1989}, traffic signs \cite{Ciresan2012}. CNN finds the association model between images and classifications by learning the training set. CNN consists of an input and an output layer, as well as multiple convolutional layers, pooling layers, fully connected layers. When classifying a target, multiple convolutional layers and pooling layers are required and they are arranged alternately. Each neuron in the convolution layer is connected to its input locally, and each connection is assigned a weight value. The output value of the neuron is calculated by the weighted sum of the corresponding local inputs plus the biased value.
\begin{figure}
\centerline{\includegraphics[width=\linewidth]{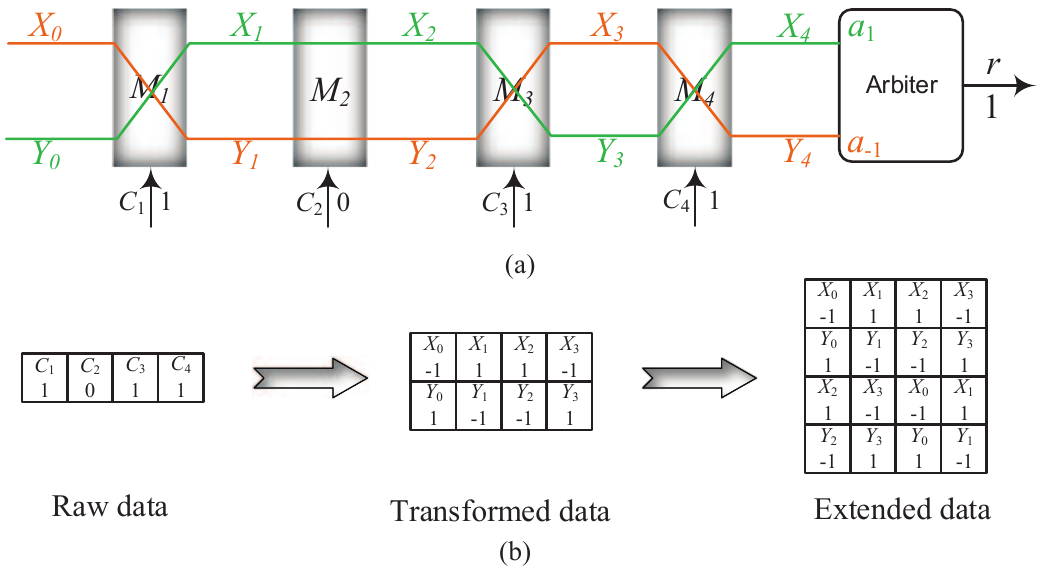}}
\caption{CRPs data transformation and extension}
\label{2}
\end{figure}

The advantage of CNN lies in the automatic extraction of features from the original pixel to the final classification, which helps to model PUF without understanding the characteristics (delays) of the PUF. In order to model PUF with CNN, we need to transform and extend the original training data (CRPs). Taking a 4-stage Arbiter PUF for example (see Fig. \ref{2}a)), when the original challenge $\emph{C}_1\emph{C}_2\emph{C}_3\emph{C}_4$ = 1011, the response is 1. In the process of data transformation, the challenge $\emph{C}_1 \sim \emph{C}_4$ are from right to left, `1' indicates that the signal is on the path to $a_1$, and `-1' indicates the path to the signal on $a_{-1}$. Therefore, in the training data, we use the path indicator (1 or -1) instead of the challenge (0 or 1). It should be noted that $\emph X_4$ is always on the path to $a_1$ and $\emph Y_4$ is always on the path to $a_{-1}$. As shown in Fig. \ref{2}(b), the dimension of transformed data is changed from 1$\times$4 to 2$\times$4. Transformed data can make CNN easier to find the mapping relationship between the challenge and the path to establish the PUF model. However, since the transformed data do not have a spatial association like the image pixels, extracting them directly using a convolutional layer may lose characteristics. Therefore, we need to further expand the transformed data. As shown in Fig. \ref{2}(b), the 2$\times$4 transformed data is further expanded to the 4$\times$4 extended data. Additionally, unlike CNN for image recognition, CNN for modeling PUF is not applicable to convolved data after a compression in pooling layers. Besides, since the pixel value of the PUF is -1 or 1, we also need to adjust the sigmoid layer to facilitate the processing of convolved data. In this way, CNN can model the PUF instance more easily.

\begin{enumerate}
  \item [4)]\emph{Artificial Neural Network (ANN)}
\end{enumerate}

ANN is a self-adaptive learning system composed of interconnected computing nodes called neurons. A strong motivation to use ANN is given by the universal approximation theorem: a two-layer neural network containing a limited number of hidden neurons can fit any function with high accuracy \cite{Rosenblatt1957}.

The simplest neural network consists of a layer with several neurons, called the single layer perceptron (SLP) \cite{Rosenblatt1957}. For each neuron, all input vectors are weighted, added, biased and applied to an activation function to generate an output. In the SLP training process, the neuron updates its weight and bias based on the linear feedback function of the training set prediction error. When prediction accuracy of the trained model reaches the preset termination condition, e.g., the preset number of iterations, the training process will be stopped. SLP can only solve the problem which is linearly separable, and non-linear problems require multi-layer ANNs. In our experiments, we use a 3-layer ANN to model the PUF and the activation function is sigmoid, where the first layer has 35 nodes, the second layer has 25 nodes, and the third layer has 25 nodes. In addition, the loss function of ANNs is adjusted by the RProp iterative method due to its fast convergence speed. The core parameters adjusted to build an accurate ANNs are the number of layers, the number of neurons in each layer, the activation function and the optimization algorithm.

\begin{enumerate}
  \item [5)]\emph{Evolutionary Strategy (ES)}
\end{enumerate}

ES \cite{Hansen2016} is inspired by genetics and evolutionary theory. ES is to generate different children randomly through the parent, and retain the best performing child as the parent of the next generation and then keep the cycle going. As the next generation inherits the best genes of the previous generations, species continue to evolve.

Since a PUF instance can be represented by the delay vector $\vec {\omega}$, the goal of modeling PUF with ES is to find the parameter vector $\vec {\omega}$ as accurate as possible to simulate the real PUF instance. The key idea is to generate different PUF instances randomly and keep the most suitable PUF instance as the parent of the next generation. Such process will be repeated until the child PUF instance closest to the real PUF is generated. In the next generation, child vectors usually use most of the parent's delay vectors and adopt a few randomly mutated vectors. In the ES algorithm, the most typical mutation method is to add a random Gaussian variable $N(0, \sigma)$ to each PUF instance.

The modeling accuracy is used to select the most suitable child PUF instance in this paper, and the child with the highest modeling accuracy rate will be considered most appropriate. Specifically, assuming that $\emph{R}$ and $\emph{R}$' are the $\emph{l}$ responses generated by the physical PUF and the child PUF instance when inputting the same challenges, respectively. The modeling accuracy $\emph{A}$ can be obtained by calculating the average HD between the two binary strings $\emph{R}$ and $\emph{R}$':

\begin{equation}
A = HD(R',R)/l
\end{equation}

There are many variants of the ES algorithm. The main differences between them are: 1) the number of parents kept in each generation; 2) the way that children derive from the parents; 3) the way to control the random mutation rate $\sigma$. There are two general approaches to control $\sigma$. One is to reduce the mutation rate $\sigma$ deterministically, and the other is to make the $\sigma$ adjusted adaptively according to the current execution performance of the evolutionary algorithm. In this paper, we use the covariance matrix adaptive ES (CMA-ES) to evaluate our proposed RSO-based PUF and use the default parameters in \cite{Hansen2016}. CMA-ES employs a reorganization approach where a child instance relies on multiple parent instances. CMA-ES also uses the self-adaptation, i.e., the mutation rate $\sigma$ is not controlled deterministically but adapts itself automatically depending on how the ES algorithm performs. CMA-ES has better performance than original ES in modeling PUFs.

\section{The Proposed Set-based Obfuscation}
\begin{figure}
\centerline{\includegraphics[width=\linewidth]{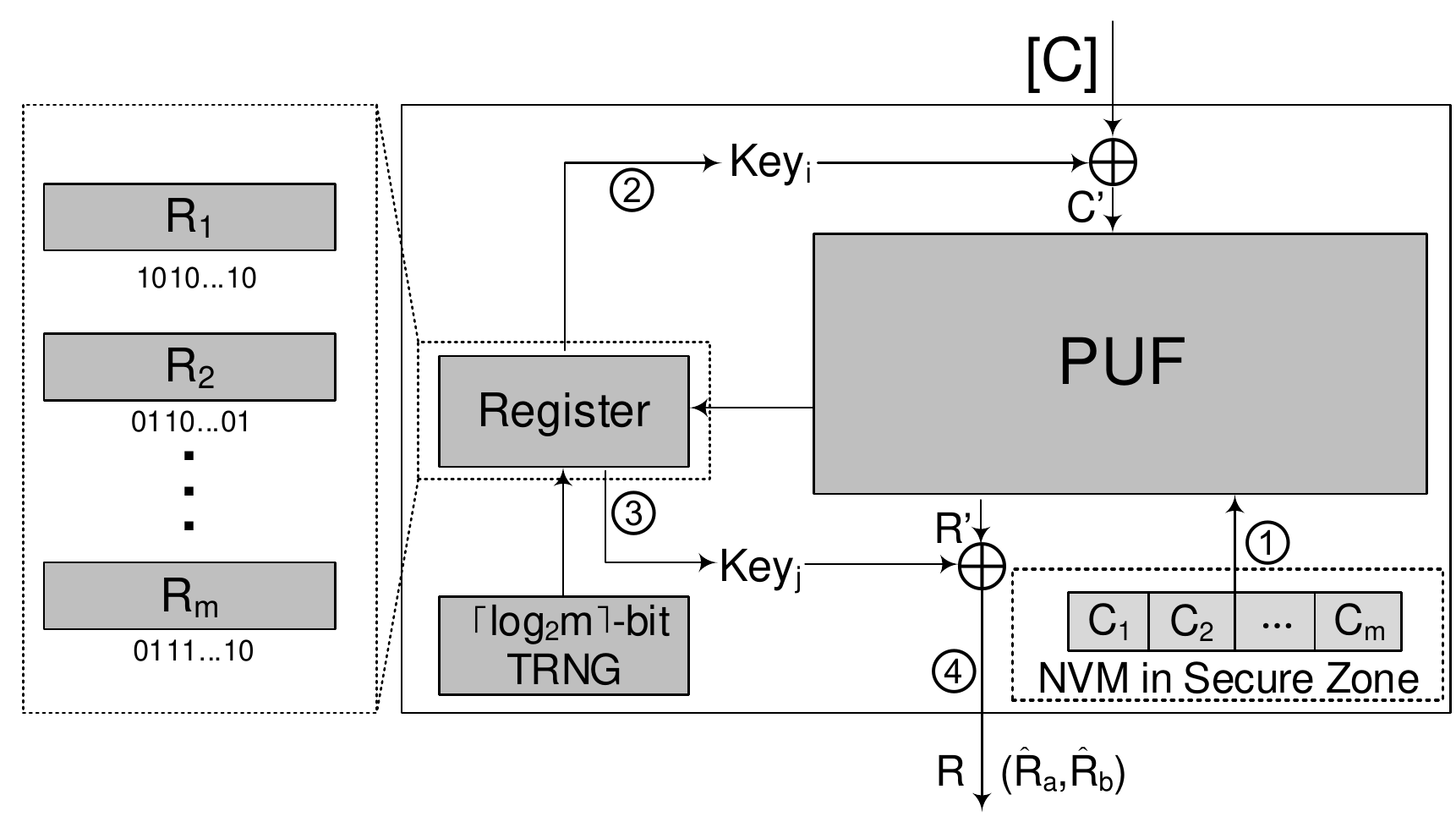}}
\caption{The structure of random set-based obfuscation}
\label{3}
\end{figure}

The modeling accuracy is related to the number of CRPs collected by attackers and the complexity of the mapping relationship of the challenge and response. Generally, the more complex the mapping relationship is and the less the number of CRPs collected, the lower the modeling accuracy will be. Our goal is to complicate the mapping relationship of the challenge and response and limit the number of CRPs collected by attackers to resist modeling attacks. This paper proposes a random set-based obfuscation (RSO) structure that uses the PUF's own multiple stable responses generated as the obfuscation source, and the set-updating mechanism will update the $set$ when the number of CRPs collected by the attacker reaches the preset threshold. In this way, ML attacks can be prevented efficiently. As shown in Fig. \ref{3}, the RSO structure consists of the XOR logic, TRNG, non-volatile memory (NVM) and registers. The concrete working principle is as follows.

First, some stable CRPs for the PUF are collected by testing, then $m\times n$ CRPs are selected from these stable CRPs, and their challenges are stored in the non-volatile memory in secure zone with directly memory access \cite{wang2018}. Second, the stored challenges are input to the PUF circuit one by one, and the generated responses are used as the obfuscation $set$ which will be temporarily stored in registers. Third, the TRNG will select two keys randomly from the registers to XOR with the challenge and response of the PUF, respectively. Finally, the obfuscated response will be sent to the server for authentication.

In the RSO, both the challenges and response are obfuscated to get the best ability to resist machine leaning. In addition, since challenges and response are obfuscated with the random stable response by bitwise XOR operation, the PUF stability will not be reduced. Therefore, the RSO can not only resist ML attacks effectively, but also address the issue that the reliability of PUFs is declined in structural non-linearization methods and CRP obfuscation methods. The process of RSO-based PUF for device authentication includes the device-side obfuscation and the server-side authentication.
\subsection{Device-side Obfuscation}
Device-side obfuscation includes the preparation phase and the obfuscation phase.
\subsubsection{\textbf{Preparation phase}}
After the PUF is manufactured, $m\times n$ stable CRPs $CR_{ob}$ = $\{C_{ob},R_{ob}\}$ = \{$[(C_{11}, C_{12}, ..., C_{1n}], R_1)$, $([C_{21}, C_{22}, ..., C_{2n}], R_2 )$, ..., $([C_{m1}, C_{m2}, ..., C_{mn}], R_m )$\} are selected during testing, where $C_{ob}$ contains $m$ challenges and all challenges and $R_{ob}$ are $n$-bit. $C_{ob}$ is stored in the NVM on the chip and $CR_{ob}$ is stored on the server.

\subsubsection{\textbf{Obfuscation phase}}
The server generates a random challenge set $ [\emph C]$ $(C_1,C_2,...,C_n)$, at the same time, the server calculates all possible responses by the parametric PUF model and the stored $CR_{ob}$, and all the responses generated will be split into two parts $R_a$ and $R_b$. Then, the server sends the challenges and all $R_a$ to the device for authentication. The RSO selects a key $\emph Key_i$  from the registers according to a random number generated by the TRNG. Then the set $\emph {C'}$ generated by XORing $ [\emph C]$ with $\emph {Key}_i$ is input to the PUF circuit one by one as the real challenges to generate the response $\emph {R'}$. Finally, another key $\emph {Key}_j$ will be selected from the registers randomly to perform XOR operation with $\emph {R'}$, and the new generated $\hat{R}$ which is split  into two parts $\hat{R_a}$ and $\hat{R_b}$ will be generated. If there is a $R_a$ that matches with $\hat{R_a}$, the $\hat{R_b}$ will be sent to the server for authentication. In this process, the real challenge $\emph {C'}$ can be expressed as:

\begin{equation}
C' = Key_i\oplus [C] = f([C_i])\oplus [C]
\end{equation}

The response $\emph {R'}$ and $\hat{R}$ can be expressed as:
\begin{equation}
R' = f(C') = f(f([C_i])\oplus [C])
\end{equation}
\begin{equation}
\hat{R} = Key_j\oplus R' = f([C_j])\oplus f(f([C_i])\oplus [C])
\end{equation}

When $\hat{R_b}$ is sent back to the server, the server will verify whether there is a $R_b$ matching with it. If it exists, the device is legal, otherwise, the device is illegal.

For example, for a 64$\times$64 RSO-based PUF ($n$ = 64), assume that the number of obfuscated keys in the set $K$ is 8 and the server sends the challenge $[C]$ = [1010...11, 1110...00, ... , 0110...10], to the PUF chip for authentication. If the TRNG chooses $Key_i$ = 0110...01 and $Key_j$  = 1010...10 to XOR with the challenge and response respectively. At this moment, $C$ will be XORed with $Key_i$ to generate $C'$ =  [1010...11, 1110...00, ... , 0110...10] $\oplus$ (0110...01) = [1100...10, 1000...01, ... , 0000...11] which are input to the PUF circuit to generate $R'$ = $f(C' ) = f([1100...10, 1000...01, ..., 0000...11])$. Then the response $\hat{R} = (Key_j\oplus R') = ((1010...10) \oplus f([1100...10, 1000...01, ... , 0000...11])) $ is generated by XORing $R'$ with $Key_j$ and $\hat{R}$ will be split into $\hat{R_a}$ and $\hat{R_b}$ for authentication. Finally, in the device side, if the $\hat{R_a}$ passes authentication, the $\hat{R_b}$ will be sent to the server for authentication.

\subsection{Server-side Authentication}
\begin{figure}
\centerline{\includegraphics[width=\linewidth]{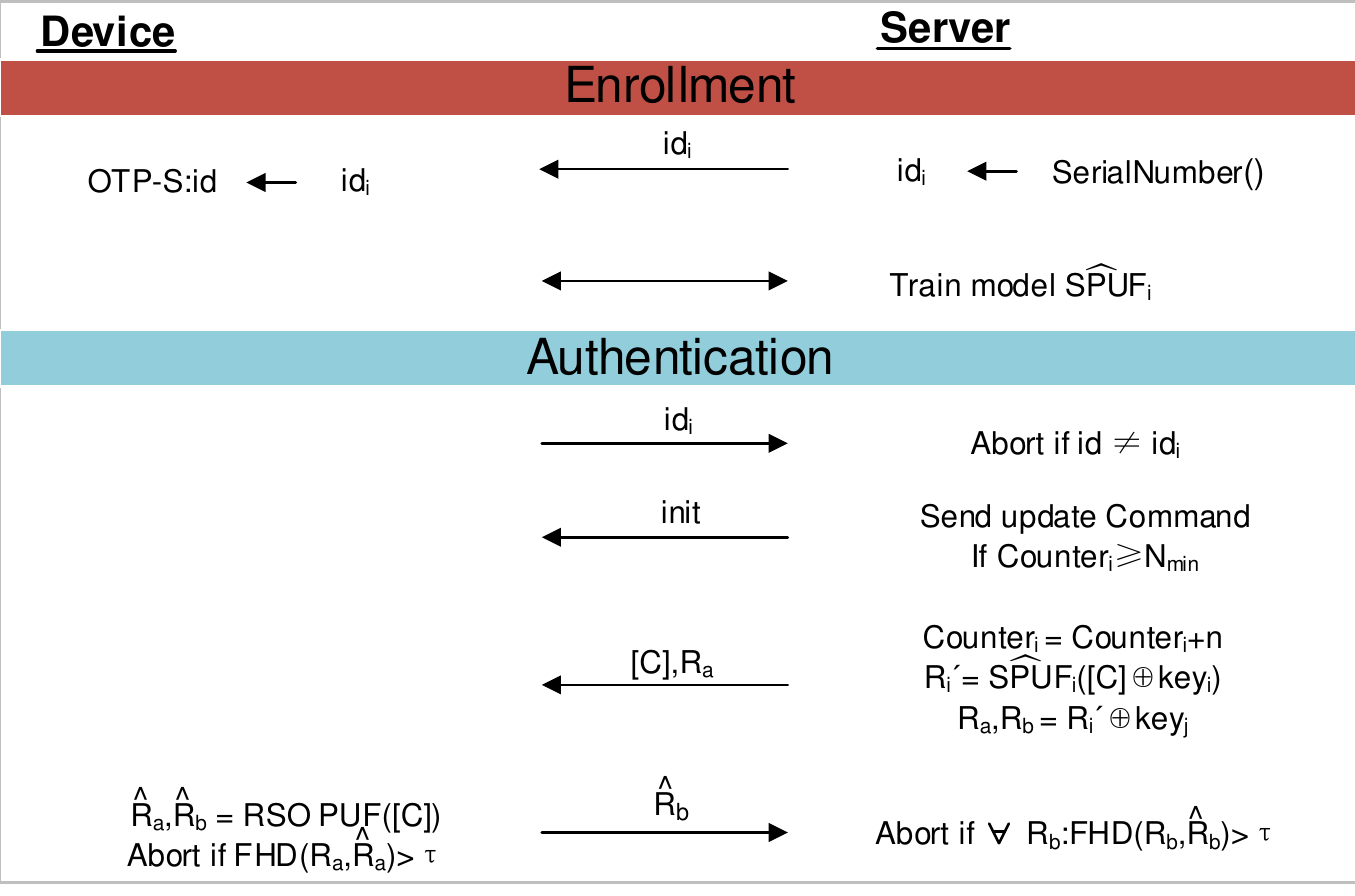}}
\caption{Protocol of RSO-based PUF for device authentication}
\label{4}
\end{figure}

On the server side, we use the PUF parametric model $\widehat{SPUF}_i$ to generate $R_a$ and $R_b$ for matching authentication. Compared with the traditional authentication method that stores all CRPs in database \cite{Suh2007,Maes2013}, the use of parametric model can reduce storage overhead greatly and improve the efficiency of server authentication. The whole authentication process on the server side is shown in Fig. \ref{4}.

\subsubsection{\textbf{Enrollment phase}}
For a $Device_i$, the device identifier $id_i$ is stored on the one-time programmable storage (OPT-S) through the e-fuse technology. At the same time, we build an accurate PUF parametric model $\widehat{SPUF}_i$ with the original CRPs and store the model parameters on the server securely.

\subsubsection{\textbf{Authentication phase}}
First, the device identifier $id_i$ will be sent from the device to the server. Second, the server will compare the sizes of $Counter_i$ and $N_{min,\epsilon}$ which is the minimum number of CRPs needed to build a model with an error rate $\epsilon$. If the value of $Counter_i$ reaches the threshold $N_{min,\epsilon}$, the server sends a key update command to the device ($ID$ = $id_i$) to update the key set $K$ on the PUF chip. When the server issues a deterministic challenge set $[C]$ and obtains $m^2$ responses which will be split into two parts $R_a$ and $R_b$, the server sends the $[C]$ and $m^2$ $R_a$ to the device according to the authentication record and updates the CRP counter $Counter_i$. In this way, it can not only prevent attackers from using the already used CRPs to conduct replay attacks \cite{Maes2013}, but also can update the keys to enhance the ability to resist ML attacks according to the recorded $Counter_i$.


In each authentication event, first, when the server receives the device identifier $id_i$, it will generate a unused challenge set $[C]$ and compute the $m$ $C'$ by XORing $[C]$ with the $m$ keys, respectively. Second, the $m$ $C'$ will be input sequentially to the parametric model $\widehat{SPUF}_i$ to generate $m$ $R'_i$. Then the $m^2$ $R$ will be generated by XORing $R'_i$ with the $m$ keys respectively and all $m^2$ $R$ will be split into two parts $R_1$ and $R_2$. After that, the server will send the challenge set $[C]$ and $m_2$ $R_a$ to the device. On the device side, $[C]$ is input to the PUF chip to generate an $n$-bit obfuscated response $\hat{R}$ which will be split into two parts $\hat{R_a}$ and $\hat{R_b}$, too. In the device side, if there is FHD$(\hat{R_a},R_a)\leq \tau$, the $\hat{R_b}$ will be sent to the server for authentication. At last, the server will compare each $R_b$ with the received $\hat{R_b}$. For the $m^2$ possible $R_b$ that may be generated, if there is FHD$(\hat{R_b},R_b)\leq \tau$, the authentication is successful, otherwise it fails.


\subsection{Set-updating Mechanism}
We propose a set-updating mechanism for RSO to prevent against all potential ML attacks. In this mechanism, when the number of CRPs collected by attackers reaches the minimum number of CRPs required for attackers to build a PUF model with an error rate $\epsilon$, the server will update the set $K$. Based on theoretical considerations (dimension of the feature space, Vapnik-Chervonenkis dimension), it is suggested in \cite{R¨¹hrmair2010} that the minimal number of CRPs which is necessary to model a $N$-stage delay based Arbiter PUF with a misclassification rate $\epsilon$ can be expressed as:

\begin{equation}
N_{min,\epsilon}^{Arbiter}\approx\frac{n+1}{2\epsilon}
\end{equation}

According to Eqn. (18), to model a 64-stage Arbiter PUF with the prediction accuracy 95$\%$, the minimum number of CRPs required for attacks is $N_{CRP,0.05}^{Arbiter}\approx650$. However, for the RSO-based Arbiter PUF, when $m$ keys are set for obfuscation, $m^2$ CRPs can be generated for each authentication. Therefore, if attackers want to build a model for the RSO Arbiter PUF with an error rate $\epsilon$, the minimum number of CRPs needed to be collected can be expressed as:

\begin{equation}
N_{min,\epsilon}^{RSO}\approx m^2\times N_{min,\epsilon}^{Arbiter}\approx m^2\times\frac{n+1}{2\epsilon}
\end{equation}

In this case, the probability that the attacker extracts valid $N_{min,\epsilon}^{Arbiter}$ CRPs from $N_{min,\epsilon}^{RSO}$ CRPs will be :

\begin{equation}
P_{ext} = \frac{m^2}{C_{N_{min,\epsilon}^{RSO}}^{N_{min,\epsilon}^{Arbiter}}}
\end{equation}

According to Eqn. (20), for a 64-stage RSO-based Arbiter PUF, when $m$ = 2 and $\epsilon$ = 5$\%$, $N_{min,\epsilon}^{RSO}$ is 2600 which is much bigger than 650 (the minimum number of CRPs required for attackers to build the model for the original 64-stage Arbiter PUF). The probability that the attacker extracts valid 650 CRPs from 2600 CRPs is $10^{-630}$. In addition, when the number of CRPs collected by the attacker reaches $N_{min,\epsilon}^{RSO}$, the server will send the key update command to the device to update the set $K$ on the PUF chip and server simultaneously. In this way, RSO-based PUF can resist potential ML attacks.

\subsection{Authentication Threshold}
The server matches the response $\hat{R}$ generated by the PUF chip and the response $R$ generated by the parametric model $\widehat{SPUF}_i$ to judge whether the authentication passes or not by setting a threshold. The minimum threshold that the response $\hat{R}$ is authenticated successfully can be defined as:

\begin{equation}
\tau = \frac{n_{tolerance}}{n}
\end{equation}
where $n_{tolerance}$ denotes the maximum number of bit-flips allowed by the PUF response when the server matches, and the response whose number of bit-flips are not greater than $n_{tolerance}$ can be authenticated successfully.

If $\tau$ is set to be greater than $n_{tolerance}/n$, the authentication speed of the PUF responses will be faster. If $\tau$ is set to be less than $n_{tolerance}/n$, the authentication speed will be slow. Therefore, the server can set $\tau$ flexibly to meet the requirements of application scenarios such as the authentication time and the security level. If the application scenario has the high requirement on the authentication efficiency, the $\tau$ can be set to be greater than $n_{tolerance}/n$; if the application scenario focuses on the security level, the $\tau$ can be set less than $n_{tolerance}/n$.

The probability of successfully authenticating a legal PUF-embedded device can be estimated as:
\begin{equation}
P_{suc} = \sum_{i = 0}^{n_{tolerance}}\left(\begin{array}{cc}n\\i\end{array}\right)\times(1-\hat p_{intra})^{n-i}\times(\hat p_{intra})^i
\end{equation}
where $\hat p_{intra}$ is the binomial probability estimator of intra-HD distributions of the strong PUF. In order to clone a PUF successfully, the prediction accuracy of the cloned PUF model should be higher than $1 - \hat p_{intra}$. For a 64$\times$64 Arbiter PUF, $P_{suc}$ is about 99.9$\%$ when $n_{tolerance}$ = 10 and $\hat p_{intra}$ = 5$\%$ which is measured in the worst case for the Arbiter PUF \cite{Lim2005}. Therefore, legitimate devices have an extremely high probability of passing authentication by setting a reasonable threshold $\tau$.

\subsection{Security Analysis}
\subsubsection{Brute force attacks}
To build a model with the accuracy $1 - \epsilon$, attackers need to create a new model based on the previous models for all possible responses generated by the challenge. Therefore, the number of models that attackers need to build to pass the authentication can be estimated by:
\begin{equation}
N_{model} = {(m)}^{2N_{min,\epsilon}^{Arbiter}}
\end{equation}
For example, when two random numbers are selected form the $set$ to obfuscate, the number of models that attackers need to build will be as high as $4^{650}$ to model a 64-stage Arbiter PUF with the error rate 5$\%$. Therefore, it is impossible for attackers to clone the RSO-based PUF by brute-force attacks.

\subsubsection{Replacement attacks}
In the proposed RSO, TRNG is used to select two random numbers from the set $K$ to XOR with $[C]$ and $R$, respectively. Therefore, for the challenge set $[C]$, RSO-based PUF may generate $m^2$$R$. However, if attackers replace all challenges in NVM with the identical challenge, then the TRNG will always choose the same key to participate in obfuscation, which will form a fixed mapping relationship between $[C]$ and $R$. In this case, the obfuscation ability of RSO will be reduced. Therefore, to prevent the replacement attack, the challenge set are stored into nonvolatile memory located in secure zone with directly memory access \cite{wang2018}.


\subsubsection{Probing attacks}
The metal wires of the RSO used to generate the delay to determine the response can be attached to the upper and lower paths of the strong PUF. Therefore, once the attacker physically detects the internal structure of the RSO-based PUF, the response generated by the PUF would be changed \cite{Gassend2008} and the entire RSO-based PUF structure would be destroyed.


\begin{figure}
  \centering
  \includegraphics[width=0.9\linewidth]{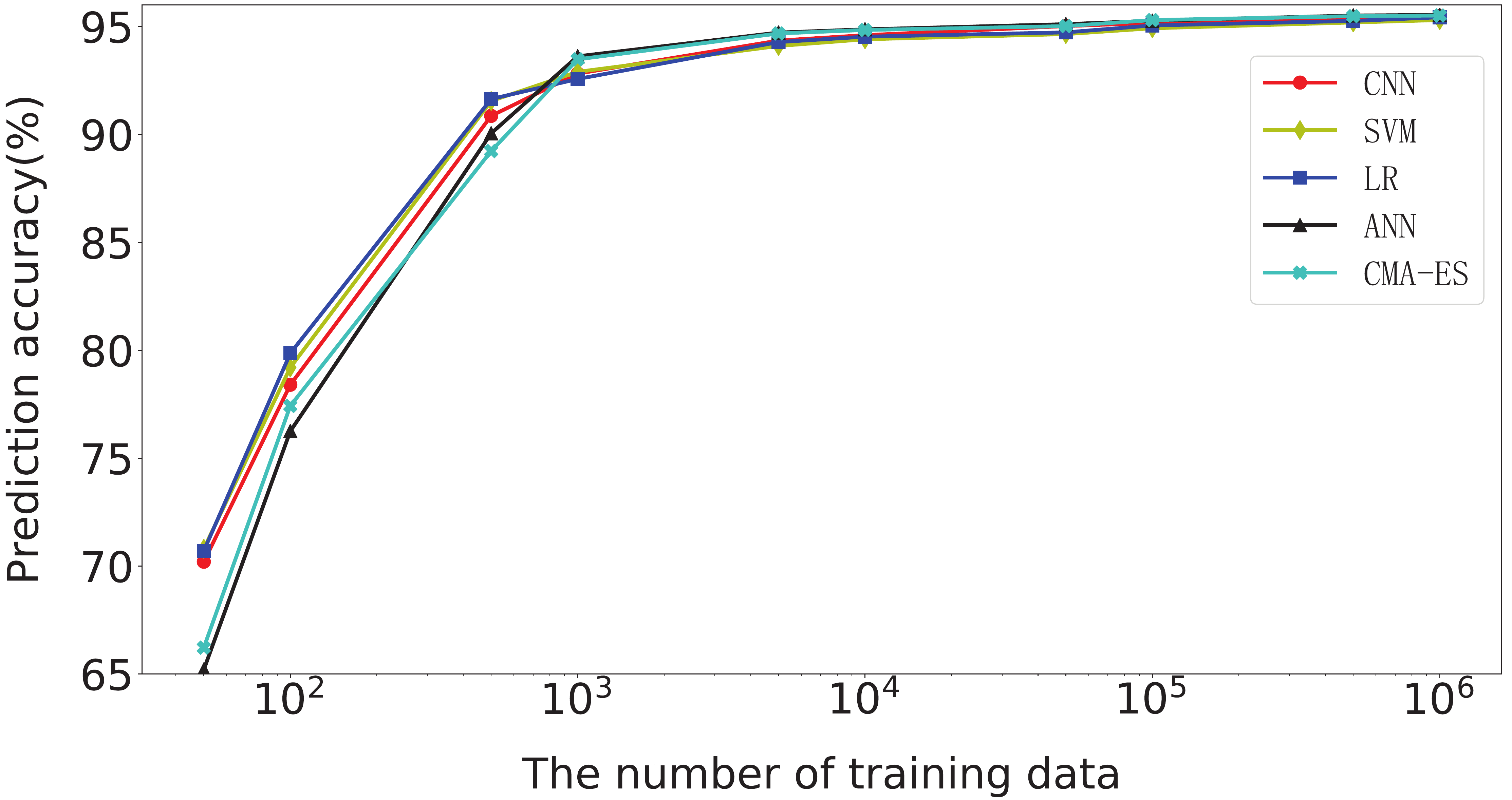}
  \caption{Modeling accuracies on the $64 \times 64$ Arbiter PUF using 1 million CRPs}
  \label{5}
\end{figure}

\begin{figure}
  \centering
  \includegraphics[width=0.9\linewidth]{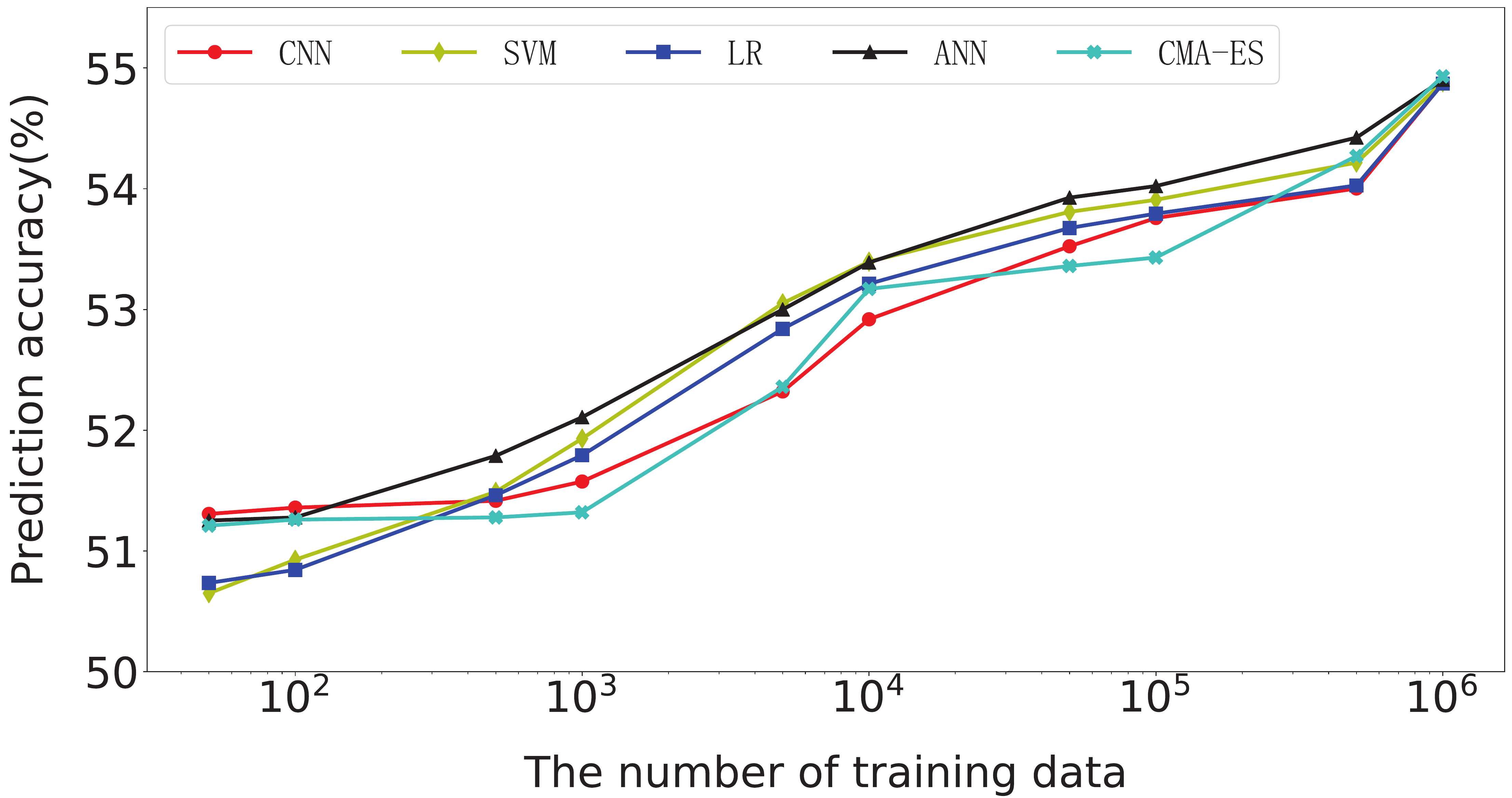}
  \caption{Modeling accuracies on the $64 \times 64$ RSO-based Arbiter PUF with 8 keys using 1 million CRPs}
  \label{6}
\end{figure}

\section{Experimental results and analysis}
\subsection{Experimental Setup}
The proposed RSO is a universal obfuscation architecture that can be used for any strong PUFs, we choose the Arbiter PUF to evaluate its resistance to ML attacks. We have collected 1 million CRPs for Arbiter PUF and RSO-based Arbiter PUF on a Xilinx Artix-7 FPGA. All experiments are conducted on the Intel i5-7400 CPU@3.0 GHz, GTX1050 GPU and 8G memory.

\subsection{ Resistance to ML attacks}
\begin{figure}
  \centering
  \includegraphics[width=0.9\linewidth]{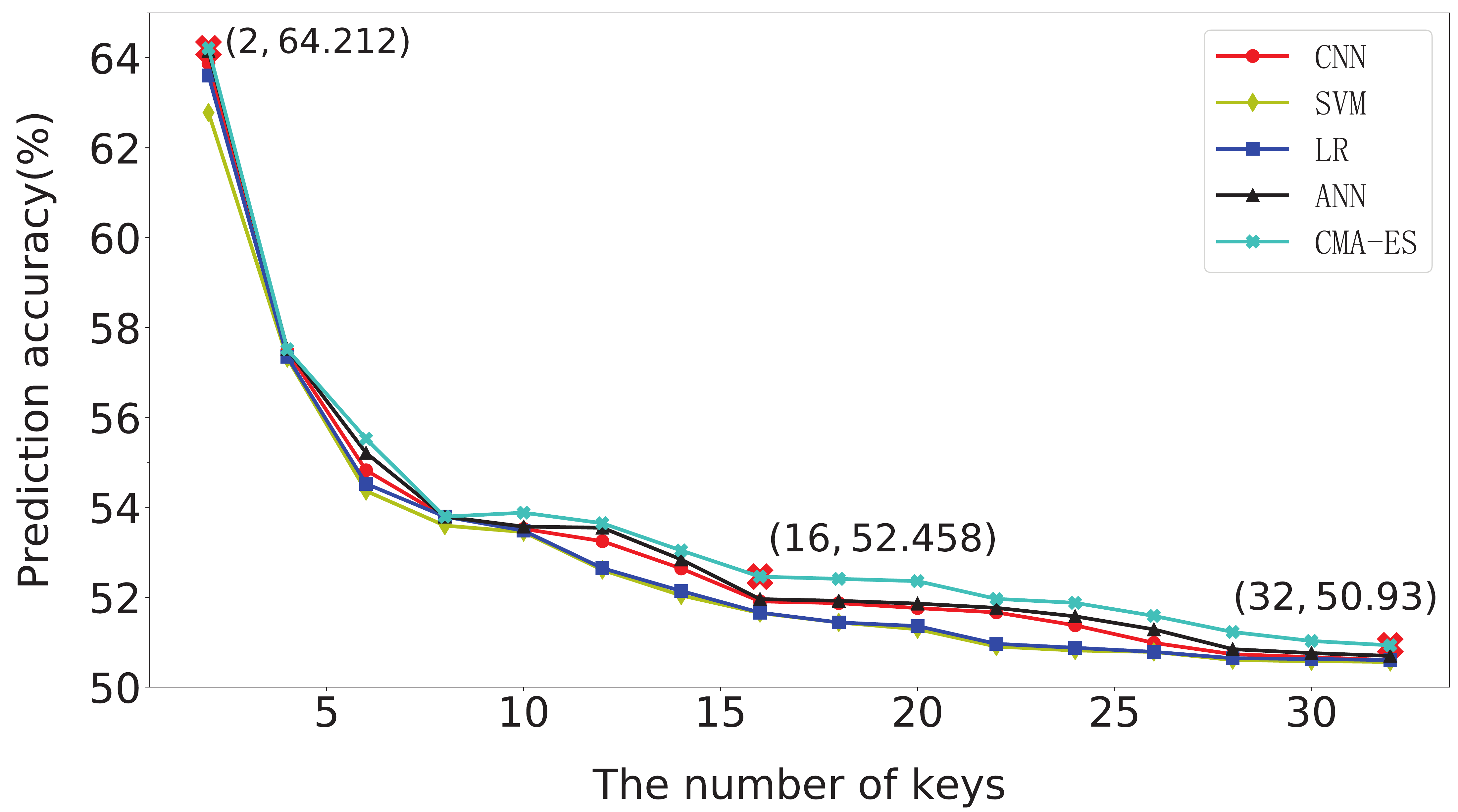}
  \caption{Modeling accuracies on the $64 \times 64$ RSO-based Arbiter PUF with different numbers of keys using 1 million CRPs}
  \label{7}
\end{figure}

We model the original Arbiter PUF and RSO-based Arbiter PUF with five ML methods, LR, SVM, ANN, CNN, and CMA-ES. In the experiments, we use the LR with the iterative function Rprop \cite{Riedmiller1993}, the SVM with the kernel function RBF \cite{Informatik1998}, a 3-layer ANNs, a CNN containing two convolution layers and two connection layers and the CMA-ES whose the fitness function is average Hamming distance to conduct ML attacks. In the model training, we divide the CRP data set into the training set (70$\%$), validation set (20$\%$) and test set (10$\%$) randomly. All trained models will be tested by 10,000 unused CRPs. The experimental results are shown in figures \ref{5}, \ref{6}, \ref{7}.

As shown in Fig. \ref{5}, without any protection strategies for the PUF, the modeling accuracy of the five ML attacks can reach 95$\%$ when 50,000 CRPs are collected, which shows that when a small number of valid CRPs are collected, attackers can clone the Arbiter PUF successfully because the average bit flip rate for a 64-stage Arbiter PUF is about 4.8$\%$. However, when the RSO is deployed for the Arbiter PUF, attackers are difficult to clone it. For example, as shown in Fig. \ref{6}, when the number of keys in the set $K$ is set to 8, even if the $K$ is not updated dynamically, compared with the original Arbiter PUF, the modeling accuracies of the five ML methods are reduced significantly. We have tested the effectiveness of the five ML algorithms on a 64-stage RSO-based Arbiter PUF. Experimental results show that even if 1 million CRPs are collected, the accuracies are lower than 55$\%$.

\begin{figure}
  \centering
  \includegraphics[width=0.9\linewidth]{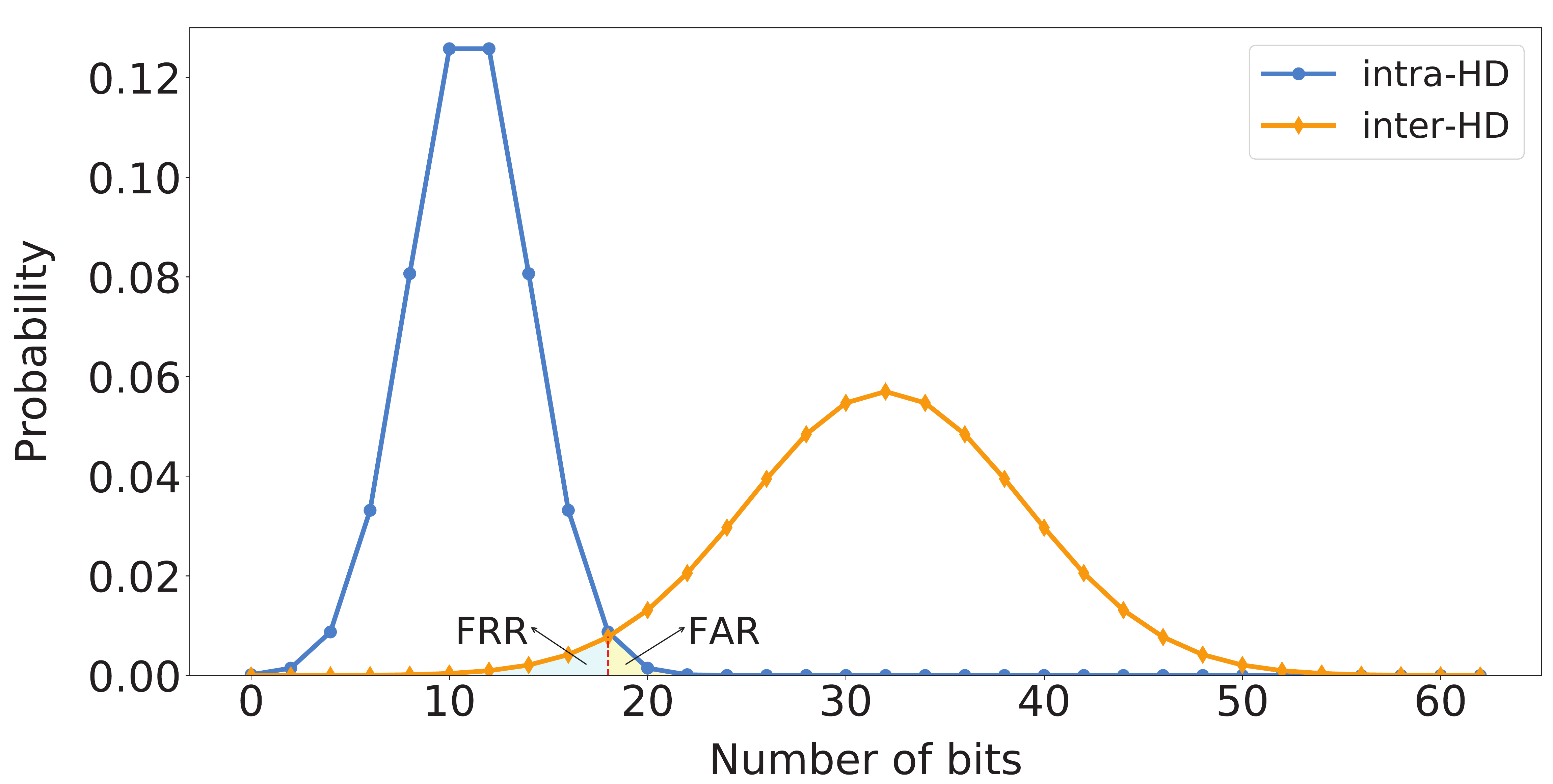}
  \caption{Distributions of intra-HD and inter-HD for the 64-bit responses}
  \label{8}
\end{figure}

We have evaluated the modeling accuracies of five ML methods on the $64 \times 64$ RSO-based Arbiter PUF with different numbers of keys using 1 million CRPs. Fig. \ref{7} shows that the modeling accuracies of LR, SVM, ANN, CNN and CMA-ES decrease significantly with the number of keys increasing. For example, when the number of keys is 2, the modeling accuracies are lower than 65$\%$; when the number of keys is 8, the modeling accuracies are lower than 60$\%$; when the number of keys is 16, the modeling accuracies are lower than 55$\%$. When the number of keys is 32, the modeling accuracies are close to 50$\%$ which is equivalent to the random guessing. Therefore, RSO-based PUF is able to resist ML attacks effectively.

\begin{table*}[!htbp]
\caption{Modeling accuracy of LR algorithm on Arbiter PUFs and RSO-based Arbiter PUF with 32, 64 and 128 stages}
\centering
\label{table2}
\begin{tabular}{*{7}{c}}
\toprule
\thead{\textbf{PUF} \\ \textbf{category}} & \thead{\textbf{Number of} \\ \textbf{challenge bits}}&\thead{\textbf{Number of} \\ \textbf{response bits}}&\thead{\textbf{Number of CRPs} \\ \textbf{in training Set}}&\thead{\textbf{Number of CRPs} \\ \textbf{in testing set
}}&\thead{\textbf{Number of} \\ \textbf{ keys in the $set$}}&\thead{\textbf{Prediction} \\ \textbf{accuracy}}\\
\midrule
\multirow{3}*{Original Arbiter PUF} & 32 &32 &1 $\times 10^4$ & $1 \times 10^4$ & N/A & 95.0$\%$\\
                      {}& 64 &64 &1 $\times 10^4$ & $1 \times 10^4$ & N/A & 95.2$\%$\\
                      {}&128 &128 &1 $\times 10^4$ & $1 \times 10^4$ & N/A & 94.8$\%$\\
\midrule
\multirow{9}*{RSO-based Arbiter PUF} & {}&{}&{}&{}&{2}&{63.61$\%$}\\
                      {}& 32 &32 &$1 \times 10^5$ & $1 \times 10^4$ & 4 &57.35$\%$\\
                      {}&{}&{}&{}&{}&{8}&{52.82$\%$}\\
                      \cline{2-7}
                      {}&{}&{}&{}&{}&{2}&{63.61$\%$}\\
                      {}& 64 &64 &1 $\times 10^5$ & $1 \times 10^4$ & 4 & 57.34$\%$\\
                      {}&{}&{}&{}&{}&{8}&{53.79$\%$}\\
                      \cline{2-7}
                      {}&{}&{}&{}&{}&{2}&{61.93$\%$}\\
                      {}&128 &128 &1 $\times 10^5$ & $1 \times 10^4$ & 4 & 55.93$\%$\\
                      {}&{}&{}&{}&{}&{8}&{52.56$\%$}\\
\bottomrule
\end{tabular}
\end{table*}

In addition, since LR performs best in terms of modeling time and modeling accuracy in five modeling attack methods, we use LR to model the 64-stage RSO-based Arbiter PUF with 32, 64 and 128-stage to evaluate the effectiveness of resistance to ML attacks. The experimental results are shown in Table \ref{table2}. For the original Arbiter PUFs, the modeling accuracy is about 95$\%$ when collecting 10,000 CRPs. However, for the RSO-based Arbiter PUF, the modeling accuracy is lower than 65$\%$ when the number of keys is 2 and 100,000 CRPs are collected; the modeling accuracy is lower than 60$\%$ when the number of keys is 4; the modeling accuracy is lower than 55$\%$ when the number of keys is 8. Therefore, the RSO-based Arbiter PUF can resist ML attacks efficiently.

Finally, RSO can update the obfuscation $set$ to greatly improve the ability against modeling attacks. Once the number of CRPs collected by attackers reaches the threshold $N_{min,\epsilon}^{RSO}$, the server will send a command to update the $set$ on the PUF chip and server synchronously. With the set-updating, the RSO can resist all advanced ML attacks in theory.

\subsection{Authentication Capability}
Fig. \ref{8} shows an example of the estimated inter-HD and intra-HD distribution of a 64 $\times$ 64 Arbiter PUF's responses. The process by which we computed these estimators guarantees that the assumed binomial distributions provide an accurate estimation, in particular for the right tail of the intra-HD distribution and for the left tail of the inter-HD distribution, because these two tails describe two undesirable errors in an authentication application: the false acceptance rate (FAR) and the false rejection rate (FRR). The authentication capability of PUF can be evaluated with the FAR and the FRR. Given an $n$-bit response, FAR denotes the probability of incorrectly accept an unauthorized device. Obviously, high FAR will bring security vulnerabilities in authentication. False rejection rate (FRR) denotes the probability of rejecting an authorized device. High FRR would result in low successful authentication rate for authorized devices. When a $n$-bit response is used for authentication, we can conduct a quantitative analysis for the FAR and FRR which are determined by the uniqueness, reliability and $n_{tolerance}$ \cite{Maes2013,Lim2005}.

\begin{enumerate}
  \item [1)]\emph{Uniqueness and FAR}
\end{enumerate}

Uniqueness is used to evaluate the difference in responses generated by different PUFs when inputting the same challenge. The paper evaluates the PUF uniqueness with the average Hamming distances:

\begin{equation}
U = \frac{2}{s(s+1)}\sum_{u = 1}^{s-1}\sum_{v = u+1}^{s}FHD(R_x,R_y)
\end{equation}
where $s$ represents the number of PUF instances, $R_x$ and $R_y$ are two $n$-bit responses generated by two PUF instances $u$ and $v$ when inputting the same challenge. The ideal value for uniqueness is 50$\%$.

For a $n\times n$ RSO-based Arbiter PUF, FAR can be expressed as \cite{Lim2005}:
\begin{equation}
FAR(n_{tolerance})=\sum_{i=0}^{n_{tolerance}}\left(\begin{array}{cc}n\\i\end{array}\right)(\hat p_{inter})^i(1-\hat p_{inter})^{n-i}
\end{equation}
where $n_{tolerance}$ is the number of flip-flop bits allowed in a response when the server matches. $\hat p_{inter}$ denotes the probability of $R_x\neq R_y$. Since the RSO-based Arbiter PUF uses the multiple challenges to generate $n$-bit response from one Arbiter PUF, the probability of  $R_x\neq R_y$ is actually the rate of different bits in two responses. Therefore, $\hat p_{inter}$ is equal to the uniqueness of the RSO-based Arbiter PUF.

We selected 10$^4$ challenges randomly to evaluate the RSO-based Arbiter PUFs with 32, 64 and 128 stages. As shown in Table \ref{table3}, the uniquenesses ($\hat p_{inter}$) are close to the ideal value 50$\%$.

\begin{enumerate}
  \item [2)]\emph{Reliability and FRR}
\end{enumerate}

Reliability is used to evaluate the stability of PUF responses generated by the same challenge in different environments. Ideally, PUF responses should remain the same under same challenges over multiple observations. Actually, a variety of environmental conditions, such as temperature, voltage and aging, may result in the delay differences in the PUF circuit and cause responses to vary. Since RSO-based Arbiter PUF obfuscates the challenge and response with the stable responses in the $set$ by bitwise XOR operation, the PUF reliability will not be reduced.

Assume that the response $R_x$ and $R_y$ are generated by the same PUF instance in different environments, and the bit-flip rate due to environmental varies is $\hat p_{intra}$. For an $n$-stage RSO-based Arbiter PUF, the FAR is the probability of FHD$(R_x,R_y)>\tau$. While the probability of FHD$(R_x,R_y)\leq\tau$ is:
\begin{equation}
P(FHD(R_x,R_y)\leq\tau) = \sum_{i=0}^{n_{tolerance}}\left(\begin{array}{cc}n\\i\end{array}\right)(1 - \hat p_{intra})^{n-i}(\hat p_{intra})^i
\end{equation}
Therefore, the FRR can be defined as \cite{Lim2005}:
\begin{equation}
\begin{split}
FRR(n_{tolerance}) = &1 - P(FHD(R_x,R_y)\leq\tau)\\
&=1-\sum_{i=0}^{n_{tolerance}}\left(\begin{array}{cc}n\\i\end{array}\right)(1-\hat p_{intra})^{n-i}(\hat p_{intra})^i
\end{split}
\end{equation}

\begin{enumerate}
  \item [3)]\emph{Analysis of Authentication Ability}
\end{enumerate}

FRR decreases with the increasing of $n_{tolerance}$, while FAR increases with the increasing of $n_{tolerance}$. High FAR or FRR is undesirable for device authentication. Therefore, we hope that the FAR and FRR can maintain balance. Assume there is a $n_{tolerance}$ that makes the FAR and FRR equal, we call this error rate as equal error rate (EER). In this case, $n_{tolerance}$ can be denoted by $n_{EER}$. However, for discrete distributions, there may not be a value that makes FAR and FRR exactly equal. Therefore,  $n_{EER}$ and EER \cite{Maes2013} can be defined as follows.

\begin{equation}
n_{EER}=argmin\{max\{FAR(n_{tolerance}),FRR(n_{tolerance})\}\}
\end{equation}
\begin{equation}
EER=max\{FAR(n_{EER}),FRR(n_{EER})\}
\end{equation}

In the experiment, $n_{EER}$ and EER are computed with the 32, 64 and 128-stage, respectively. $\hat p_{inter}$ and $\hat p_{intra}$  are measured by the RSO-based Arbiter PUF data.

\begin{table*}[!htbp]
\caption{Performance of RSO-based Arbiter PUF}
\centering
\label{table3}
\begin{tabular}{|c|c|c|c|c|c|c|c|}
\hline
\textbf{\#\emph{row}} & \textbf{$\hat {\emph{P}}_{\emph{inter}}$} & \textbf{$\hat {\emph{P}}_{\emph{intra}}$} &
\textbf{\emph{n}} & \textbf{$\emph{n}_{\emph{EER}}$} &  \textbf{\emph{FAR}} & \textbf{\emph{FRR}} & \textbf{EER} \\
\hline
\thead{\#1} & \thead{50.1$\%$} & \thead{5.0$\%$} & \thead{32} & \thead{6} & \thead{2.6$\times10^{-4}$} &  \thead{2.6$\times10^{-4}$}&\thead{8.7$\times10^{-4}$}\\
\hline
\thead{\#2} & \thead{49.8$\%$} & \thead{4.8$\%$} & \thead{64} & \thead{13} & \thead{1.1$\times10^{-6}$} &  \thead{1.7$\times10^{-6}$}&\thead{1.7$\times10^{-6}$}\\
\hline
\thead{\#3} & \thead{49.7$\%$} & \thead{5.2$\%$} & \thead{128} & \thead{27} & \thead{2.3$\times10^{-11}$} &  \thead{ 8.9$\times10^{-11}$}& \thead{ 8.9$\times10^{-11}$}\\
\hline
\end{tabular}
\end{table*}

As shown in Table \ref{table3}, for the 32$\times$32 RSO-based Arbiter PUF, FAR is closest to FRR when $n_{EER}$ = 6. In this case, EER is 8.7 $\times$ $10^{-4}$, which is higher than the standard of 10$^{-6}$ (the required identification performance of an identification system is determined by its application, but for most practical applications a FAR and FRR both $\leq10^{-6}$, and hence an EER $\leq10^{-6}$, is minimally desired \cite{Maes2013}). For the 64$\times$64 RSO-based Arbiter PUF, FAR is closest to FRR when $n_{EER}$ = 13. In this case, EER is 1.7 $\times$10$^{-6}$, which meets the standard in practical applications. For the 128$\times$128 RSO-based Arbiter PUF, FAR is closest to FRR when $n_{EER}$ = 27. In this case, the EER value is 8.9$\times$$10^{-11}$ which can be applied in practice well.

\section{Comparisons}
In this section, we compare the RSO-based PUF with the Controlled-PUF \cite{Gassend2008}, PUF-FSM \cite{Gao2017} and Slender PUF \cite{Rostami2014} to evaluate hardware overhead and security.

\begin{figure}
  \centering
  \includegraphics[width=\linewidth]{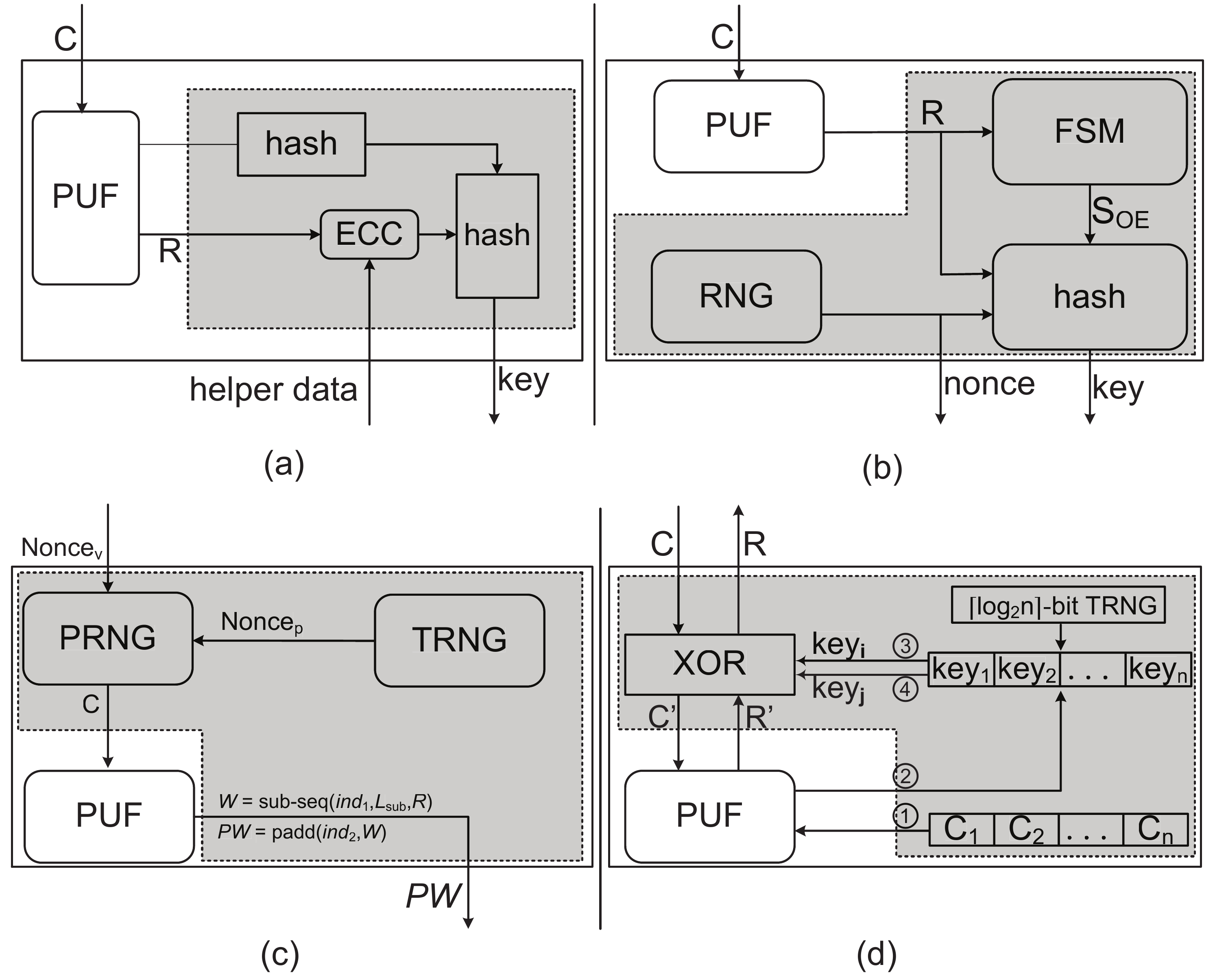}
  \caption{(a) Controlled PUF structure \cite{Gassend2008}. (b) PUF-FSM structure \cite{Gao2017}. (c) Slender PUF structure \cite{Rostami2014}. (d) RSO-based PUF structure}
  \label{9}
\end{figure}

\subsection{Hardware Overhead Comparison}
As shown in Fig. \ref{9}(a), the Controlled PUF \cite{Gassend2008} adds the two hash circuits to obfuscate both the challenge and response, which requires an error correction code (ECC) to correct the unstable PUF responses. However, the hash circuit will incur high hardware overhead, and the ECC unit is also expensive and the hardware overhead is exponentially related to the number of error correction bits, which make the PUF difficult to be applied to resource-constrained devices.

As shown in Fig. \ref{9}(b), based on the Controlled PUF, the PUF-FSM \cite{Gao2017} removes the hash circuit on the challenge side and replaces the ECC unit with the FSM state conversion structure at the response side to reduce the hardware overhead. However, the hash circuit on the response side still consumes considerable hardware resources. Additionally, the PUF-FSM protocol requires transferring more than 160$\times$64 = 10240 challenge bits \cite{Delvaux2017}, which is more expensive than storing or transferring the helper data of a fuzzy extractor \cite{Herrewege2012}.

As shown in Fig. \ref{9}(c), the hardware overhead of Slender PUF \cite{Rostami2014} consists of 4 parallel 128-stage Arbiter PUFs, TRNG, FIFO, LFSR and control logic. The TRNG in PUF chip generates a nonce $nonce_p$ first. Then, combining with the nonce $nonce_v$ received from the server, a random seed is generated by concatenating $nonce_p$ and $nonce_v$. The generated seed is used by a pseudo-random number generator (PRNG) to output the challenge $C$ which will be input to the PUF. At last, Slender PUF will select a sub-sequence $W$ of the response randomly and pad it with a random binary string to create a bitstream $PW$ of the response length for authentication. However, $4$ parallel Arbiter PUFs and related circuits still consume considerable hardware overhead.

As shown in Fig. \ref{9}(d), the hardware overhead of RSO includes the XOR logic, TRNG and NVM. XOR logic for obfuscation and HD detection consumes 75 LUTs and 46 DFFs. TRNG is used to select a key from set $K$ randomly to obfuscate the challenge and response. Many TRNG have been implemented on FPGAs \cite{Majzoobi2011,Yang2018}. For example, the TRNG \cite{Yang2018} only consumes 10 LUTs and 5 flip-flops on the Xilinx Spartan-6 FPGA chip. Nevertheless, the additional hardware overhead incurred by TRNG can be avoided in practical application: 1) the TRNG has been used in many systems and hence can be reused; 2) the metastable PUF responses in Arbiter PUFs can be used as the random number \cite{Ranasinghe2005}.

\begin{table}
\caption{Hardware Overhead Comparison with methods \cite{Gassend2008,Gao2017,Rostami2014}}
\centering
\label{table4}
\begin{tabular}{|c|c|c|c|}

\hline
\thead{\textbf{Type}}&\thead{\textbf{\#LUT}}&\thead{\textbf{\#DFF}}&\thead{\textbf{RAM}}\\
\hline
\thead{Controlled PUF \cite{Gassend2008}} & \thead{1830} & \thead{3020} & \thead{N/A} \\
\hline
\thead{Slender PUF\cite{Rostami2014}} & \thead{1168} & \thead{1400} & \thead{4KB} \\
\hline
\thead{PUF-FSM \cite{Gao2017}} & \thead{960} & \thead{1500} & \thead{N/A} \\
\hline
\thead{RSO-based PUF} & \thead{405} & \thead{181} & \thead{4KB} \\
\hline

\end{tabular}
\end{table}

In order to further demonstrate the low overhead of our proposed RSO, we compare the RSO with recent obfuscation methods \cite{Gassend2008}, \cite{Gao2017}, \cite{Rostami2014}. Based on a 128-stage Arbiter PUF implemented on the Xilinx Artix-7 FPGA chips, the resources (LUT, DFF and RAM) consumed by these different structures are summarized in Table \ref{table4}. A 128-stage RSO-based Arbiter PUF consumes only 395 LUTs and 176 flip-flops, the TRNG consumes 10 LUTs and 5 flip-flops, the required NVM is about $2^3\times64\times$64 bits = 4KB when $m$ = 8. Therefore, the hardware overhead of RSO is much smaller than other obfuscation structures.

\section{Conclusion}
In this paper, we propose a new obfuscation technique for strong PUFs, named random set-based obfuscation (RSO), where a true random number generator is used to select any two random numbers from the set which is derived from the strong PUF to obfuscate challenges and responses respectively with the XOR operation to prevent attackers from collecting effective CRPs to perform ML attacks. With the set-updating, ML attacks become more harder. Experimental results demonstrate its advantages of strong resistance to ML attacks and low hardware overhead.

\vfill

\end{document}